
\input jnl
\input reforder
\input psfig
\ignoreuncited
\hskip 4truein UCSB-TH-91-41
\smallskip

\title            QUANTUM GRAVITY AND STRING THEORY--
                  WHAT HAVE WE LEARNED?$^*$
\vskip 1truein
\author           Andrew Strominger

\affil            Department of Physics, University of California
                  Santa Barbara, CA 93106
                  Bitnet: andy@voodoo

\vskip 4.25truein
\hsize=6.5truein
\noindent
$\overline{^*~To}~appear~in~the~proceedings~of~the~Sixth~Marcel~Grossman~Conference~on$
\noindent$~General~Relativity,~June~24{-}29~(1991),Kyoto,~Japan.~~~~~~~~~~~~~~~~~~~~~~$
\endtitlepage
The last time I had the opportunity to speak at a Marcel Grossman meeting
was six years ago in Rome, just following the exciting realization
[\cite{GS.2}] that string theory might actually describe our universe. Since
that time there has been an enormous amount of activity in the subject.
Much of the progress has been rather technical and difficult to understand
by those outside of the field. In this talk I am going
to give a very non-technical distillation of some of this work. This
is not meant to be a review. Rather I will
focus on several results of the past six years
with qualitative implications for the problem of
quantum gravity, and begin with some perspective on that problem.

\noindent{\bf The Problem of Quantum Gravity}

Quantum mechanics and general relativity are perhaps the two greatest
achievements of twentieth-century physics. However the two theories
are at odds: the standard recipe (quantum field theory) for quantizing
a classical field theory fails when applied to general relativity.
New ultraviolet divergences appear at every order in perturbation theory,
requiring an infinite number of counterterms for renormalization. This is
unacceptable: even if one could stomach a theory with an infinite number
of free parameters, it is doubtful that the resulting theory is stable and
unitary at the Planck scale.

This conflict is extremely fortunate for physicists.
A glance at the history of physics, depicted in Figure 1,
reveals that all
great leaps in our understanding of the universe originated
in such contradictions. For example, the ultraviolet catastrophe which
occurs in the thermodynamics of classical electromagnetism led to the
development of quantum mechanics.
\goodbreak
\topinsert
\centerline{\psfig{figure=qg.fig1}}
\endinsert
The last problem in this figure, quantum gravity, was
noted in the fifties by Pauli. It took
three decades to arrive at superstring theory, the first--and only--plausible
mathematical resolution of this problem to date. The basic idea is that all
particles, on closer inspection (${10^{-33}}\,\,{cm})$, are actually tiny
closed loops of string. Further, all the forces and particles of
nature (photons, quarks, etc.) are simply different vibrational modes
of the same fundamental string.

Let me list the main successes of superstring theory:

\item{1.} Perturbatively reconciles quantum mechanics and general
relativity (probably, some intricacies of
higher order perturbation theory remain to be sorted out).
\item{2.} First example of a truly unified theory in which all
particles and forces are different manifestations of a single object.
\item{3.} Could in principle describe our universe.
\goodbreak
\topinsert
\centerline{\psfig{figure=qg.fig2}}
\endinsert
The main failure of superstring theory can be put in perspective by an
elaboration of Figure 1. From Figure 2 we
see that experiment played a key role in the resolution of past
contradictions. Unfortunately, because of the enormous energies
involved, it is difficult to conceive of an experiment which will aid in
the reconciliation of quantum mechanics and general relativity. While
no stone should be left unturned, it seems likely that we will have to live
with this state of affairs for the forseeable future (although optimists
remain[\cite{john}]).

While this is certainly sad, it does not prevent us
from investigating the conceptual implications of string theory (assuming
it is correct). One expects the successful reconciliation of quantum
mechanics and general relativity to profoundly affect our view of the
universe. Perhaps even if string theory is not physically correct
there are useful lessons to be learned from such investigations.

In the following three subsections
I will describe three such tantalizing themes which have
recurred in investigations of string theory. These are

\item{I)} Duality and the existence of a fundamental shortest length,

\item{II)} Infinite numbers of local symmetries,

\item{III)} Quantum hair on black holes.

\item{I.} {\bf Duality/Minimal Length}

In colliding particles at energy ${E}$ one probes distances of order
$$
{\Delta}{X} {\sim} {{{\hbar}}\over{E}}.
$$
\noindent in units where ${c}={1}$.
This is not quite true for colliding
strings because strings are extended objects.
When thrown together at high energies they have a tendency to ``squash out''
to a size of order ${G_NE}$, where ${G_N}$ is Newton's constant.
This has been carefully analyzed
in [\cite{YGMACV}].
The distance one can really probe is therefore roughly
$$
{\Delta}{X}{\sim}{{\hbar}\over{E}} + {G_NE},
$$
\noindent and
very short distances cannot be probed by going to very high energies.
The minimum distance one can probe in any scattering experiment is in fact
$$
{\Delta}{X_{min}} {\sim} {\sqrt{G_N\hbar}} = {L_p},
$$
\noindent where ${L_p}$ is
the Planck length. To a physicist what can't be measured
(even in principle) does not exist, therefore we should suspect that
lengths less than ${L_p}$ simply do not exist in string theory. This
might also explain how string theory cures the ultraviolet divergences
of quantum gravity.

This view gains support from some fascinating work on a lattice formulation
of string theory. In reference [\cite{kleb}], a regulated version of
bosonic string theory is defined in which the string is composed of
discrete bits joining neighboring points of a spacetime lattice.
A certain measure is defined for summing over all string configurations
and the theory is  studied as a function of the string tension, $T$ (but
the spacetime lattice spacing $a$ is not varied). Astonishingly,
it is found that for
a range of values of $T$ (measured in lattice units) the
spectrum of this lattice theory agrees exactly with that of continuum
bosonic string theory, even for finite $a$. Thus
an exact desciption of bosonic string theory can be obtained {\it without}
any
short distance degrees of freedom!

These observations tie in neatly with
another well-known phenomenon in string
theory: duality. To understand duality in string theory, let us first
consider ordinary Kaluza-Klein compactification of ${d} = {10}$ gravity
on a 6-torus of radius ${R}$. At distances much greater than ${R}$,
the universe is effectively four-dimensional. It may be described by
the effective action
$$
{S_{\rm eff}} = {{1}\over{G_N}}{\int}{d^4}{x}{\sqrt{-g}} (
R-
\sum_{i=1}^\infty \biggr(({\nabla\phi_i})^2
+ ({m^{kk}_i})^2{\phi_i^2}) + {\cdots}\biggl).
$$
\noindent In addition to several massless fields, there is an infinite tower
of massive Kaluza-Klein fields, ${\phi_i}$,
which are relics of the underlying
ten-dimensional physics. The masses ${m^{kk}_i}$ of these Kaluza-Klein
fields are proportional to the eigenvalues of the spin-two operator
governing linearized fluctuations of the metric about the flat 6-torus
of radius ${R}$. Dimensional analysis then implies that the masses vary with
${R}$ as
$$
{m^{kk}_i}{\sim}{{1}\over{R}}.
$$
\noindent As ${R}{\rightarrow}{\infty}$, an infinite number of fields move
down to zero mass in an attempt to recover the continuous spectrum of
ten-dimensional flat space.

\goodbreak
\topinsert
\centerline{\psfig{figure=qg.fig3}}
\endinsert

Since string theory resembles a field theory at large distances, these
Kaluza-Klein modes also
appear in a 6-torus compactification of ten-dimensional
string theory. However in this case it is not the whole story. In
addition there are ``winding'' modes, depicted in Figure 3, corresponding
to a single string which wraps (perhaps many times) around the 6-torus. From
the four-dimensional perspective, this appears as a particle
with mass given by the length of the string $({\sim}{R})$ times the
string tension $T$, which is roughly
${L^{-2}_p}$:
$$
{m^w_i}{\sim}{R}/{L^2_p}.
$$
\noindent We see that these modes vary with ${R}$ in the opposite fashion
of the Kaluza-Klein modes: they become light at very small ${R}$ when only
a short string is required to wrap around the torus. In fact it was shown
in [\cite{GSBKYS}] that the spectrum of masses is invariant under the duality
transformation:
$$
{R}{\leftrightarrow}{L^2_p}/{R}.
$$
While this may seem rather peculiar, it is even more surprising that this
duality extends to interactions. Indeed it was shown in [\cite{STR.31}] that

{\it No physical experiment can distinguish a compactification of radius
${R}$ from one of radius ${L^2_p}/{R}$.}

Again to a physicist that which
cannot be measured does not exist, so we must conclude that a very
small torus is the same thing as a very large torus.

At this point you might raise the following objection: ``Surely if the torus
is 10 meters across, I can simply go in with my ruler and measure its
radius, thereby distinguishing it from a torus of radius ${L^2_p}/{10}\,{m}
= {10^{-71}}{m}$!'' The problem with this procedure is that there are two
kinds of rulers: those constructed from Kaluza-Klein modes and those
constructed from winding modes. There will be no invariant way to determine
which kind of ruler you have used, and accordingly whether you have actually
measured ${R}$ or ${L^2_p}/{R}$.

This duality between long and short distances is not confined to the 6-torus,
and has been discovered in a wide variety of different situations. This
leads one to believe that in some sense long and short distances should
be identified in string theory. Clearly this will require
a fundamental revision in our usual notion of a spacetime
continuum. Finding the proper notion to replace the spacetime
continuum and describing this duality is one of the exciting current
problems in string theory.

\item{II.} {\bf Infinite Symmetry}

String theory can be thought of as a field theory with an infinite
number of particles, one for each vibrational mode of the string.
The massless modes (e.g. graviton, photon) are associated with
local gauge symmetries. Since these are unified with the
massive modes by the string, one expects local symmetries associated
with the infinite tower of massive modes as well. Of course since
these modes are massive, the infinite set of local symmetries must
be spontaneously broken.

An infinite number of local symmetries might also explain, as
argued in [\cite{GRO}], the high energy behavior of scattering amplitudes.
For large center-of-mass energy ${\sqrt{s}}$, the amplitude ${A}({s})$
is exponentially suppressed:
$$
{A}({s})^s{\vec\rightarrow}^\infty{e^{-s/M_p^2}}.
$$
\noindent where ${M_p}={\sqrt{\hbar/G_N}}$ is the Planck mass.
This suppression may be due to high-energy symmetry restoration:
the S-matrix near infinite ${s}$ is constrained by an infinite number of
symmetries which force it exponentially to zero.

In ordinary field theories, spontaneously broken symmetries lead to
spontaneously broken Ward identities among the Feyman diagrams. The
Feyman diagrams of string theory (in the BRST formalism)
do indeed exhibit an infinite
number of such identities. This provides our most concrete understanding
of the infinite string symmetries.

But this is not good enough. In ordinary gauge theories, the
Ward identities can be derived beginning with a gauge-invariant
action, and the nature of the symmetries is better understood as an
invariance of this original action than as identities among Feynman diagrams.
A similar understanding is desirable for
the infinite string symmetries.

Fundamentally new ideas are probably required before such an understanding
is obtained, but some partial progress has been made in the context of ``string
field theory''. While it is generally believed for a variety of reasons
that this is ultimately the wrong direction, the partial results in this
direction nevertheless do much to clarify the nature of the problem
and possible solutions, as follows.
The infinite
number of fields in string theory can be assembled into one very large
``string field,'' an infinite multiplet, which we shall denote
${A}$. The action can then (for some string theories)
be elegantly written in the Chern-Simons
form: [\cite{WIT.6}]
$$
{S} = {\int}({A}{Q}{A} + {{2}\over{3}} {A^3})
$$
\noindent where the ${Q}$ is a (nilpotent) generalization of the
spacetime laplacian,
${\int}$ is a generalization of integration and the last term is the
interactions.
This action is invariant under the symmetry
$$
{\delta_{\epsilon}}{A} = {Q}{\epsilon} + [{\epsilon},{A}].
$$
\noindent Since ${\epsilon}$, like ${A}$, is a string field this
amounts to an infinite
number of ordinary spacetime symmetries.

This construction partially realizes the goal of representing the string
symmetries in a simple way. However, it is still not quite satisfactory
because of the homogeneous term ${Q}{\epsilon}$ in the symmetry
transformation law for ${\delta_\epsilon}{A}$. This means that the
vacuum state ${A}={0}$ spontaneously breaks all symmetries for which
${Q}{\epsilon}{\not=}{0}$, which includes the symmetries associated
to the massive string modes. This is like expanding the standard model
lagrangian about the broken symmetry minimum of the Higgs potential. The
nature of the symmetry is much more evident when the lagrangian
is expanded about the state of unbroken symmetry.

{\it Is there a state of unbroken symmetry in string theory?}
This is a fascinating
question. Some light has been shed on it by the discovery [\cite{HIKKOHLRS}]
that a redefinition of the string field ${A}$
$$
{\tilde A} = {A} - {A_0},
$$
\noindent where ${A_0}$ is a certain classical solution of ${S}$,
leads to the action
$$
{S}[{\tilde A}] = {{2}\over{3}}{\int}{\tilde A}^3
$$
\noindent and gauge transformation law
$$
{\delta_\epsilon}{\tilde A} = [{\epsilon},{A}].
$$
\noindent The vacuum ${\tilde A} = {0}$ is then the state of unbroken string
symmetry since it is left invariant by all symmetry transformations.
It is further of interest to note that, as ${Q}$ has disappeared
in this reformulation, there is no reference to spacetime geometry in
${S}[{\tilde A}]$. This suggests that the notion of spacetime appears
only as a result of spontaneous breakdown of the infinite
string symmetry!

While tantalizing, the above description is beset by possibly incurable
difficulties which are too technical to describe here. Most
notably, the state ${\tilde A} = 0$ is represented
in a singular fashion. Witten's ``topological field theory'' [\cite{WIT.2}]
attempts to find a better description of this state of
unbroken symmetry, but has so far not fully succeeded.
While it is generally believed that string theory contains an infinite number
of local symmetries, their proper description is yet to be found.

\item{III.} {\bf Quantum Hair}

Consider the action
$$
{S} = {{1}\over{G_N}}
{\int}{d^4}{x}{\sqrt{-g}} (R- {H}_{\mu\nu\lambda}{H^{\mu\nu\lambda}}
+ {\cdots})
$$
$$
{H}_{\mu\nu\lambda} = {\partial}_{[\mu}{B}_{\nu\lambda]}
$$
\noindent The axion field strength ${H}$ describes one pseudoscalar degree of
freedom $a$ defined by $H_{\mu\nu\lambda}={\epsilon_{\mu\nu\lambda}}^\rho
\partial_\rho a$. This arises as part of the low-energy effective action for
string theory, but most of the comments of this section are more general
and pertain to any action with axions and gravity. The action ${S}$
has the solution
$$
{g_{\mu\nu}} = {g^s_{\mu\nu}}
$$
$$
{H_{\mu\nu\lambda}} = 0
$$
$$
{B_{\mu\nu}} = {q}{\epsilon_{\mu\nu}}
$$
\noindent where ${g^s_{\mu\nu}}$ is the standard Schwarzchild black hole
metric  and the two-form ${\epsilon}$ is tangent to the two spheres of
constant ${r},{t}$ and normalized so that
$$
{\int}_{S^2}{B} = q
$$
\noindent for any two-sphere surrounding the black hole.

Classically, the ``hair'' ${q}$ is unobservable (since it does not
enter into the field strength) and therefore uninteresting. However,
as argued in [\cite{STR.43}], it is quantum mechanically detectable. Strings
(either fundamental or solitonic) couple to ${B_{\mu\nu}}$ via
the action
$$
{S_B} = {{T}\over{2}} {\int_\Sigma}{\epsilon^{ij}}
{\partial_i}{X^{\mu}}{\partial_j}{X^\nu}{B_{\mu\nu}}
$$
\noindent (where ${T}$ is the string tension)
which is just the integral of ${B}$ over the string worldsheet
${\Sigma}$. It follows that for a string which encircles a black
hole
$$
{S_B} = {T}{q}
$$
\noindent and ${S_B}$ is otherwise zero. One then concludes
that [\cite{STR.43}]

{\it A string thrown around a hairy black hole gets a phase
${e^{iTq/\alpha^{\prime}{\hbar}}}$. This phase is quantum mechanically
detectable through interference experiments.}

Classically, the no-hair theorems assert that the final states of a black hole
are labelled by just a few parameters, such as charge, mass and angular
momentum. Quantum mechanically we now find that additional quantum numbers
are required.
This basic notion of quantum hair on black holes has been further developed
and elaborated in a variety of contexts, most notably that of discrete
gauge theories[\cite{KWP}].

An important feature of quantum hair is that it persists even if the
axion gets a mass through spontaneous symmetry breaking\footnote{$\dagger$}
{As in the Cremmer-Scherk mechanism. It does not persist for explicit
symmetry breaking in which case domain walls will form and confine
strings.}[\cite{WABL}].

The existence of quantum hair is relevant to issues surrounding
Hawking radiation. According to Hawking, a black hole loses its
mass by radiation of a thermal spectrum of particles. This may be
described as a pair creation process in which one particle goes
down the black hole and the other escapes to infinity. The black hole
eventually loses most (or all) of its mass and
settles down to a final state classically characterized by just a
few parameters such as mass or charge. This final state can
not carry any information about what went down the hole and so, Hawking
argues, there is a net loss of information and quantum
coherence.

This argument is affected by the possibility of quantum hair. We
have just learned that quantum mechanically there is additional information
contained in the final state of the black hole. However, since an
arbitrarily large amount of information can fall into the black hole,
a qualitative effect on the question of coherence loss could occur only
in a theory with infinite varieties of quantum hair. Since quantum
hair is associated with (spontaneously broken)
local symmetries, infinite varieties of quantum hair might be expected
in a theory with infinite varieties of local symmetries. But we have just
learned that string theory may be precisely such a theory. This led
Schwarz [\cite{SCHW.2}] to the following bold
conjecture:

{\it The infinite string symmetry leads to infinite varieties
of quantum hair on black holes. This hair encodes complete information about
what went down the black hole, and no information is lost in the process
of Hawking evaporation.}

Clearly string theory is providing interesting perspectives on the many
conceptual issues arising in the reconciliation of quantum mechanics and
general relativity.

\noindent{\bf TOY STRINGS}

It used to be thought that strings only made sense in 10 or 26
spacetime
dimensions.  However, it was realized relatively recently[\cite{rob}]
that in fact strings can be made mathematical
sense of in any number of dimensions, although thay have some
unphysical properties away from the critical dimensions of 10 or 26.
Nevertheless, these so-called ``non-critical'' strings can provide
interesting theoretical laboratories for investigating questions in
quantum gravity.

In particular there has been much recent activity in the study of
string theories in two or fewer spacetime dimensions. The amazing
``matrix model''
techniques lead to a closed form solution[\cite{matrix}] for many of these
theories to all orders in quantum perturbation theory--in some
cases even non-perturbatively!

A related - and even more recent - development is the discovery[\cite{bhole}]
of an exact classical solution of non-critical string theory
corresponding to a two-dimensional black hole. This
opens the possibility of investigating the fascinating issues
surrounding the problem of Hawking radiation in the context of a
consistent quantum theory of gravity.

These are striking discoveries. I would not have suspected that
it would prove possible to obtain such complete analytic understanding
of these toy string models. However, I think that the full benefit of these
recent technical breakthroughs is yet to be reaped. Can
any light be shed on the physical questions discussed in the previous
sections? Currently there is much activity in this subject, with encouraging
progress. Perhaps I will be fortunate enough to report on this at the
seventh Marcel Grossman meeting.

\noindent{\bf  CONCLUSION}

Barring enormous luck or inspiration, string theory is unlikely
to be experimentally verified or disproved in the forseeable future
(although optimists remain [\cite{john}]).
Nevertheless it provides a rich and fascinating model to study the conceptual
revolution in our view of the universe which will inevitably
accompany the unification of quantum mechanics and general relativity.
On this issue progress is being made in leaps and bounds.
\endpage

\centerline{\bf ACKNOWLEDGEMENTS}
I would like to thank the organizers for the invitation to speak.
This work was supported in part by DOE grant DE-FG03-91ER40168.

\references

\refis {GS.2} M. B. Green and J. H. Schwarz, {\sl Phys. Lett.}
{\bf 149B}, (1984) 117.

\refis {YGMACV} T. Yoneya in ``Wandering in the Fields'' (World Scientific:
Singapore (1987));
D. J. Gross and P. F. Mende, Phys. Lett. {\bf 197B}
(1987) 129, D. Amati, M. Ciafaloni and G. Veneziano, Phys. Lett.
216B (1989) 41.

\refis {GSBKYS} M. B. Green, J. H. Schwarz and L. Brink, Nucl.
Phys. {\bf 198B} (1982) 474; K. Kikkawa and M. Yamasaki, {\sl Phys. Lett.}
{\bf 149B} (1984) 357; N. Sakai and I. Senda {\sl Prog. Theor. Phys.
Suppl.} {\bf 75} (1986) 692.

\refis {bhole} E. Witten, Phys. Rev. {\bf D44} (1991) 314; I Bars and
D. Nemeschansky,  Nucl. Phys. {\bf B348} (1991) 89, S. Elitzur, A. Forge
and E. Rabinovici, Nucl. Phys. {\bf B359} (1991) 581; G. Mandal,
A. M. Sengupta and S. R. Wadia Lett. Mod. Phys. {\bf A6} (1991) 1685.

\refis {rob} R. Myers, Phys. Lett. {\bf B199} (1988) 371,
J. Polchinski, Nucl. Phys. {\bf B324} (1989) 123.

\refis  {matrix} E. Brezin and V. Kazakov, Phys. Lett. {\bf B236} (1990)
14;
M. Douglas and S. Shenker, Nucl. Phys. {\bf B335} (1990) 635;
D. Gross and A. Migdal, Phys. Rev. Lett. {\bf 64} (1990) 127.

\refis {STR.31} V. P. Nair, A. Shapere, A. Strominger and F. Wilczek,
Nucl. Phys.
{\bf 287B}, 402 (1987).

\refis {GRO} D. J. Gross, Phys. Rev. Lett. {\bf 60} (1988) 1229.

\refis {WIT.6} E. Witten,
Nucl. Phys. {\bf 268B}, 353 (1986).

\refis {HIKKOHLRS} H. Hata, K. Itoh, T. Kugo, H. Kunitomo and K. Ogawa,
Phys. Lett. {\bf 175B}, (1986) 138; G. Horowitz, J. Lykken, R. Rohm
and A. Strominger,
Phys. Rev. Lett. {\bf 57} (1986) 283; A. Strominger, Nucl Phys.
{\bf 294B} (1987) 93.

\refis {WIT.2} E. Witten,
Comm. Math Phys. {\bf 117}, (1988) 353.

\refis {STR.43} M. Bowick, S. Giddings, J. Harvey, G. Horowitz and
A. Strominger,
Phys. Rev. Lett. {\bf 61} (1988) 2823.

\refis {KWP} L. M. Krauss and F. Wilczek, Phys. Rev. Lett. {\bf 62}
(1989) 301; J. Preskill, Caltech Preprint,
CALT-68-1671 (1990) and references therein.

\refis {WABL} F. Wilczek (private communication, unpublished); T. J. Allen,
M. J. Bowick and A. Lahiri, Phys. Lett. {\bf 237B}
(1990) 47.

\refis {SCHW.2} J. H. Schwarz, Caltech preprint, CALT-68-1728 (1991).

\refis {kleb} I. Klebanov and L. Susskind, Nucl. Phys. {\bf B309}
(1988) 175.

\refis{john}  J. H. Schwarz, Caltech preprint, CALT-68-1740, Ginsparg\#9108022
(1991).

\endreferences
\endit
\end